\documentclass[preprint2]{aastex}
\usepackage{txfonts}
\usepackage{natbib}
\usepackage{rotating}
\usepackage{subfigure}
\usepackage{multirow}
\usepackage{color}
\slugcomment{Not to appear in Nonlearned J., 45.}
\shorttitle{QBP with depth and latitude}
\shortauthors{Simoniello, R. and et al.}
\begin{document}
\title{The quasi-biennial periodicity as a window on the solar magnetic dynamo configuration}%What is the mechanism driving the quasi-biennial periodicity in the Sun?}
\author{R.~Simoniello\altaffilmark{1,2}, K.~Jain\altaffilmark{3}, S.C.~Tripathy\altaffilmark{3}, S.~Turck-Chi\`eze\altaffilmark{1}, C.~Baldner\altaffilmark{4}, W.~Finsterle\altaffilmark{2}, F.~Hill\altaffilmark{3}, M.~Roth\altaffilmark{5}}
\affil{$^{1}$Laboratoire AIM, CEA/DSM-CNRS-Universit\'e Paris Diderot; CEA, IRFU, SAp, centre de Saclay, F-91191, Gif-sur-Yvette, France\\
\email{rosaria.simoniello@cea.fr}}
\affil{$^{2}$PMOD/WRC Physikalisch-Meteorologisches Observatorium Davos-World Radiation Center, 7260 Davos Dorf, Switzerland}
%\and
%\author{K.~Jain and S.~C.~Tripathy}
\affil{$^{3}$National Solar Observatory, Tucson, AZ 85719, USA}
%\and
%\author{S.~Turck-Chi\`eze}
%\affil{Laboratoire AIM, CEA/DSM-CNRS-Universit\'e Paris Diderot; CEA, IRFU, SAp, centre de Saclay, F-91191, Gif-sur-Yvette, France}
%\author{C.~Baldner and S.~Basu}
\affil{$^{4}$W.W. Hansen Experimental Physics Laboratory, Stanford University, Stanford, CA 94305-4085, USA}
%\author{W.~Finsterle}
%\affil{PMOD/WRC Physikalisch-Meteorologisches Observatorium Davos-World Radiation Center, 7260 Davos Dorf, Switzerland }
%\author{F.~Hill}
%\affil{$^{3}$National Solar Observatory, Tucson, AZ 85719, USA}
%\and
%\author{M.~Roth}
\affil{$^{5}$Kiepenheuer Institute for Solar Physics, Freiburg}
\begin{abstract}
Manifestations  of the solar magnetic activity through periodicities of about 11 and 2 years are now clearly seen in all solar activity indices.
In this paper, we add information about the mechanism driving the 2 year period
by studying the time and latitudinal properties of
acoustic modes that
 are sensitive probes 
 of the subsurface layers. 
  We use almost 17 years of high quality resolved data provided by the Global Oscillation Network Group (GONG) to investigate the solar cycle changes in $p$-mode frequencies for  spherical degrees $\ell$ from 0 to 120 and 1600~$\mu$Hz $\le\nu\le$ 3500$~\mu$Hz. For both periodic components of solar 
  activity, 
 we locate the origin of the frequency shift in the subsurface layers and put in evidence for a sudden enhancement in amplitude just in the last few hundred kilometers. We also show that, in both cases, the size of the shift increases towards equatorial latitudes and from minimum to maximum of solar activity, but, in agreement with previous findings,
 the quasi-biennial periodicity (QBP) causes a weaker shift in mode frequencies and a slower enhancement than the one caused by the 11 year cycle.
We compare our observational findings with the features  predicted by different models that try to explain the origin of this QBP and conclude that the observed properties
could result from the beating between a dipole and quadrupole magnetic configuration of the dynamo.
\end{abstract}
\keywords{Sun: helioseismology - activity - dynamo}
%\titlerunning{Double Magnetic activity cycle}
%\authorunning{R.Simoniello}
\maketitle

\keywords{}


\section{Introduction}
Long term time series open a new perspective in the study of stellar activity cycles as they revealed the existence of a complex scenario. In fact stars can give rise to all types of periodicities from none to multiple cycles \citep{Bali95,Bra98, Mes02, Ola02,Boo07}. A beating between dipolar and quadrupolar components  of dynamo's magnetic configuration might explain the apparent multiple periods observed in some stars \citep{Mos04,Flu04,Mos08}.

The Sun also show several periodicities on different time scales longer and shorter than the sunspot cycle \citep{Bai03, Uso07, Kol09}. The quasi-biennial periodicity (QBP) has recently received a great deal of interest, as it appears in all solar activity proxies. Its amplitude is particularly strong near the solar maximum, although it doesn't seem to characterize every solar cycle \citep{Kri02,Vec08,Vec09}.
Several mechanisms have been proposed so far to explain the origin of the QBP like a second dynamo mechanism  generated by the strong rotational shear extended from the surface down to 5$\%$ below it \citep{Ben98a,Ben98b} or the instability of magnetic Rossby waves in the tachocline \citep{Zaq10,Zaq11}.

 Recently the discovery of the shortest cycle so far measured has also been reported on solar type star  \citep{Met10}. This phenomenon could be related to the mid-timescale magnetic variations recently
identified in HD 49933 from asteroseismic observations \citep{Gar10} and to the solar QBP clearly visible in  helioseismic measurements of low degree modes \citep{Bro09a, Sal10, Sim12}. 

The helioseismic changes in $p$-mode parameters 
%has already proven to b 
are strongly correlated to the cyclic behavior of solar magnetic activity. 
%as the visible manifestation of the 11 yr-cycle clearly appears in $p$-mode frequency and amplitude temporal variations. 
The mode amplitude is linked to the excitation of the mode and its changes is due to mode conversion \citep{Sim10}. The mode frequency shift results from the effect of the magnetic field on the acoustic cavity extension. The seismic signatures of the solar QBP have been interpreted as the result of a second dynamo mechanism by \cite{Fle10} and \cite{Bro12}. \cite{Jai11} and \cite{Sim12} discussed other scenarios such as the signature of the relic solar magnetic field, different dynamo modes or a separate mechanism from the main dynamo. 

With the hope to better understand the origin of the solar QBP, we investigate the temporal variation in $p$-mode frequency of low and intermediate degree modes over solar cycle 23 and ascending phase of solar cycle 24 using data provided by the Global Oscillation Network Group. We perform a detailed study of the 2 yr signal extracted by the total shifts by investigating its properties as function of mode frequency, penetration depth and latitude. When scaled by their mode inertia all the modes with different degree vary in the same way, so looking simultaneously at numerous modes within selected frequency bands should let the smaller effect of time variabilities emerging. We will afterwards confine our study in the subsurface layers selecting only modes whose lower turning point is above 0.90 solar radius and by tuning the ratio $\frac{m}{\ell}$, we will select modes sensitive to specific latitudes. The characterization of the shift with frequency, penetration depth and latitude will provide the chance to verify whether or not the origin of the shift over the 2 yr cycles could differ from the one induced over the 11 yr cycle. Finally the properties of the 2 yr signal will be also compared with the features predicted by different models and this should allow us to gain information on the mechanism driving the cyclic behaviour of solar magnetic activity.
\section{Data analysis}
\subsection{Mode frequency determination}
The {\bf G}lobal {\bf O}scillation {\bf N}etwork {\bf G}roup (GONG) provides nearly continuous and stable velocity images of the Sun since May 1995. It consists
of six instruments deployed worldwide, based on a Michelson interferometer using the Ni line at 676.8nm. 
Mode parameters (frequency, full width and amplitude) for each ($n$,$\ell$,$m$) are estimated up to $\ell$=150
by applying the standard GONG analysis \citep{And90}. The peak-fitting algorithm has two types of criteria to judge the quality of the fit to a mode \citep{Hil98} and based on these quality flags, we removed few outliers from the analysis. The 
individual $p$-mode parameters are afterwards made 
publicly available (http://gong2.nso.edu/archive/). In this work we analyze the temporal variations of $p$-mode frequencies covering almost 17 years of observations starting from May 1995 up to January 2012.
\subsection{Central frequency of the multiplets}
The spatial structure of the acoustic modes in the Sun
can be described by associated Legendre functions $P^{m}_{\ell}$(x), where $\ell$ is the degree, $m$ is the azimuthal order, running from -$\ell$ to $\ell$, x is the cosine of the colatitude $\theta$. The degeneracy 
among the modes of the same $\ell$ and $m$ is broken by
rotation and asphericity. The labelling  of a mode is completed by the radial order $n$ (number of radial intersections or harmonics). The  frequencies of a mode of specific $n$, $\ell$ multiplet are given using the following polynomial expansion:
\begin{equation}
\nu_{n, \ell, m}=\nu_{n,\ell}+\sum a_{j}(n,\ell)P_{j}^{\ell}(m)
\end{equation}
where the basic functions are polynomials related to the Clebsch-Gordan coefficients $C_{j0 \ell m}^{\ell, m}$ \citep{Rit91}by
\begin{equation}
P_{j}^{\ell}(m)=\frac{\ell\sqrt(2\ell-1)!(2\ell+j+1)!}{(2\ell)!\sqrt(2\ell+1)}C^{\ell, m}_{j,0, \ell, m}
\end{equation}
The frequency $\nu_{n\ell}$ is the so-called central
frequency of the multiplet and we will also use this parameter as it provides a measure of the global activity of the Sun.
\subsection{Frequency shift determination}
In this work we look for temporal variations in $p$-mode frequencies caused by changes in magnetic activity levels. Within this context the frequency shift can be defined as the 
difference between the frequencies of the corresponding modes observed 
at different epochs and reference values taken as average over the minimum phase of solar activity ($\delta\nu\equiv\nu_{i}-\nu(B_{0})$) or
 as the difference between the mode frequency at certain date and its temporal mean ($\delta\nu_{i}\equiv\nu_{i}(B)-\bar{\nu_{i}}$) \citep{How02}. We choose the first approach as it provides the advantage to direct compare the shift with other publications, because doesn't depend on the inclusion of new data sets.
 Since the frequency shifts has a well-known dependency on frequency and mode inertia \citep{Jai01}
, we consider only those modes that are present in all data sets and the shifts are scaled by the mode inertia \citep{Chr91}. The mean frequency shifts is calculated from the following relation:
\begin{equation}
\delta\nu(t)=\frac{\sum_{n,\ell,m}{\frac{Q_{n,\ell}}{\sigma^{2}_{n,\ell,m}}}\delta\nu(t)}{\sum_{n,\ell,m}\frac{Q_{n,\ell}}{\sigma^{2}_{n,\ell,m}}}.
\end{equation}
The weighted averages of these frequency shifts were 
then calculated in two different frequency bands: 
\begin{enumerate}
\item low frequency band 1600~$\mu$Hz$\le\nu\le$ 2500~$\mu$Hz;
\item high frequency band 2500~$\mu$Hz$\le\nu\le$ 3500~$\mu$Hz.
\end{enumerate}
This frequency dependence analysis will tell us to which depths the 2 year signal is the most sensitive. In fact the mode frequency determines the position of the upper turning point (UTP) of the waves and as the mode frequency increases as the UTP approaches the solar surface. 
Seismic observations over the 11 year cycle have already shown that, the increase in the magnitude of the shift is predominantly a subsurface phenomenon as it strongly increases in the upper few hundred kilometers \citep{Cha01}.
\section{Analysis of the QBP signatures in $p$-mode frequency shift}
\subsection{Frequency dependence}
\begin{figure}[!h]
%\begin{center}
\includegraphics[width=3.5in]{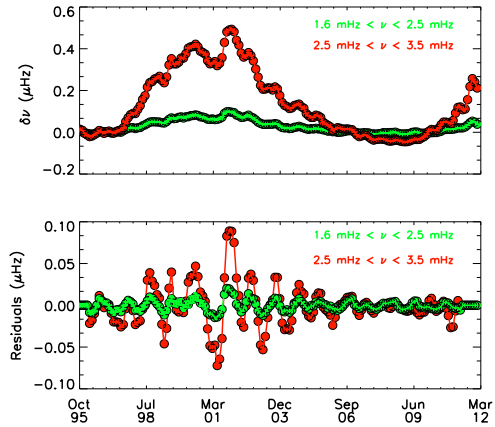}
\caption{The frequency dependence of the shift in the low (green) and high frequency band (red) over solar cycle 23 (upper panel) and the residual 2 yr cycles (lower panel).}
\label{fig:freq_dep}
%\end{center}
\end{figure}

 We investigate the solar cycle changes in $p$-mode central frequency averaged over spherical degree $\ell$=0-120. In order to extract the quasi-biennial signal from the total shifts, we subtracted the 11 year envelope by using a boxcar average of 2.5 year width \citep{Fle10}.
Fig.~\ref{fig:freq_dep} compares the temporal changes in $p$-mode frequency in the two frequency bands over solar cycle 23 (top panel) and more specifically for the 2-year period (bottom panel). The error in shifts are of the order of $10^{-3}\mu$Hz for both frequency bands. The presence of a quasi-biennial modulation is clearly visible in both frequency bands and over the whole period of observation. %The QBP signal becomes rather strong over periods coinciding with solar maximum, while it gets weaker away from the maximum. 
There are several interesting features to underline in the magnitude of the shift over the 2 year cycle  and that one can compare to previous studies on low degree modes:
\begin{itemize}
\item it is rather weak;
\item it increases by a factor of $\approx$ 3 from low to high frequency band. This enhancement is smoother compared to the stronger increase over the 11 yr cycle;
\item it becomes faint over the descending phase of solar cycle 23 and the low and high frequency band are identical in this descending phase.
\end{itemize}
The other interesting feature is that 
 over the ascending phase of solar cycle 23 the signal is larger in the high frequency band, while this does not occur over ascending phase of solar cycle 24. This might imply that the ongoing cycle 24 will be weaker compared to solar cycle 23.
\subsection{Assessing the significance}
We
investigate the significance of the QBP signal in $p$-mode central frequency averaged over spherical degree $\ell$=0-120 and
 in the two frequency bands, 
\begin{figure*}
\begin{center}
\mbox{\subfigure{\includegraphics[width=3.2in]%
{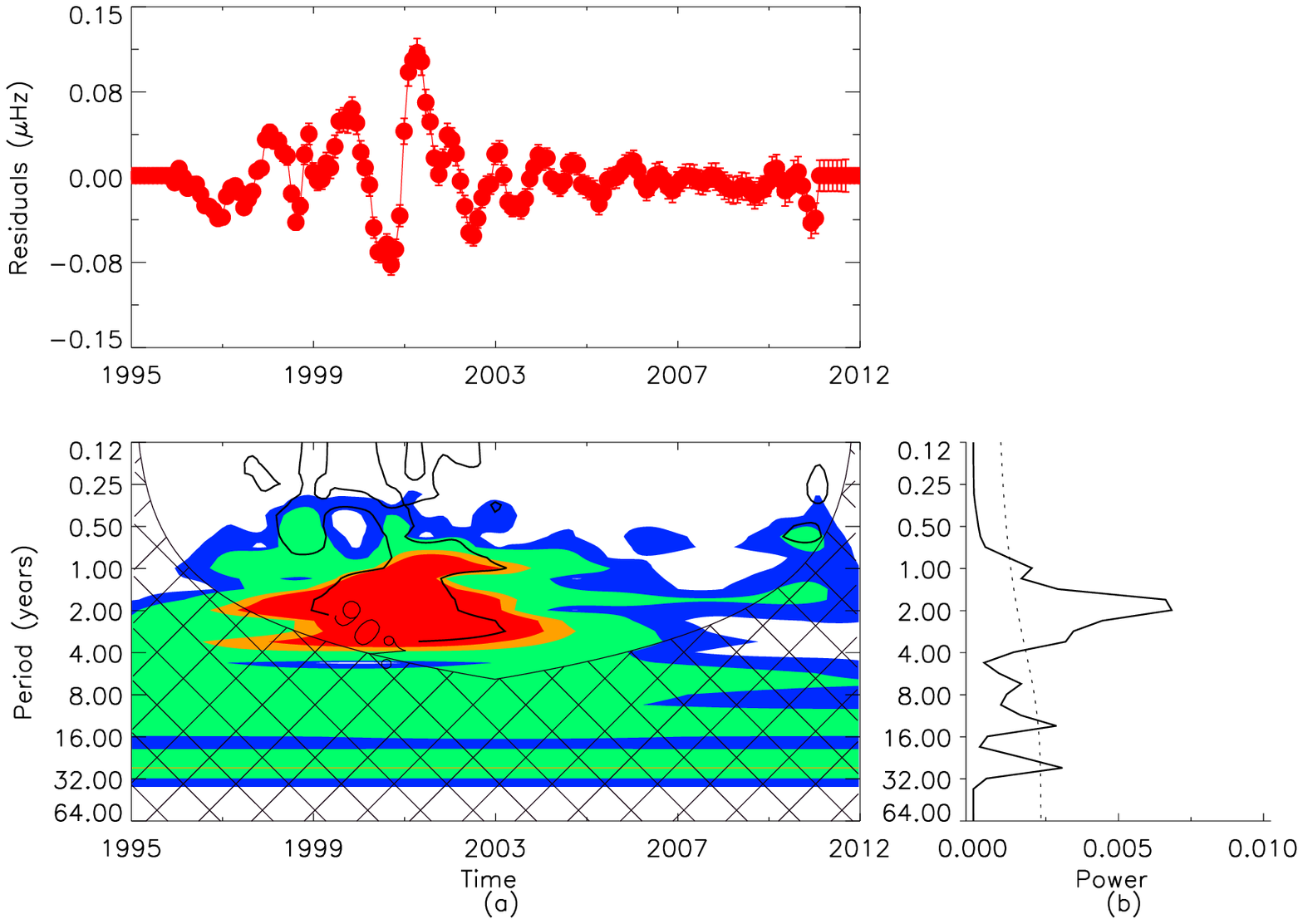}}%
\quad
\subfigure{\includegraphics[width=3.2in]{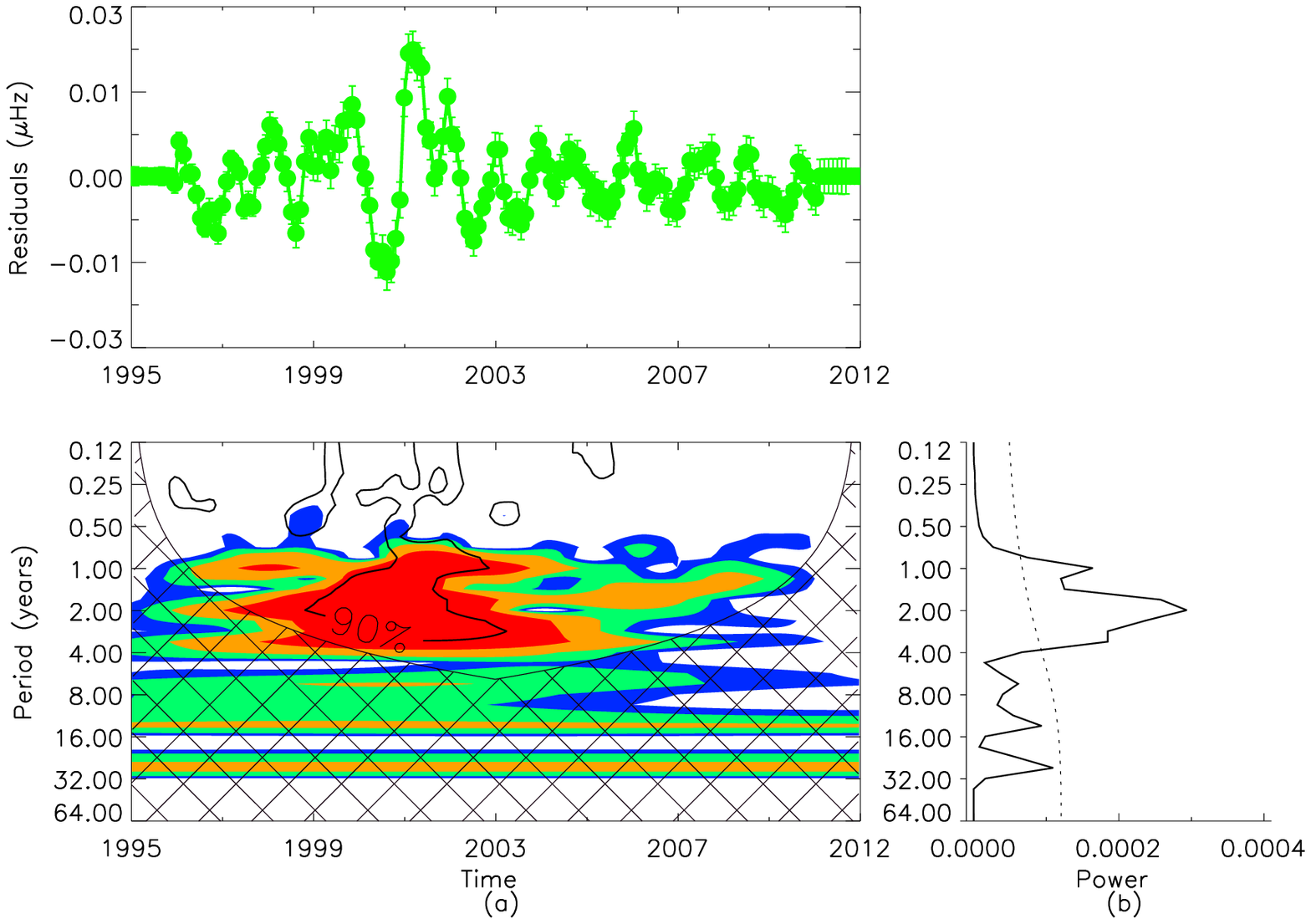}}}%\quad
%\subfigure{\includegraphics[width=2.3in]{wavelet_gong_l2_v2.ps}}
\caption{Top panels: Temporal evolution of $p$-mode frequency shifts averaged over $\ell$=0,120 spherical degree and in the high (left panel) and low (right panel) frequency bands. (a) The local wavelet power spectrum. White represent areas of little power, red those with the largest power. Solid rings identify areas at certain power level over time and only those that achieved 90$\%$ confidence level are labelled. Areas outside the cone of influence are doubtful. (b) Global wavelet power spectrum. The dotted line are the 90$\%$ of confidence level.} 
\label{fig:wavelet}
\end{center}
\end{figure*}
by applying the wavelet analysis developed by \cite{Tor98}. The upper panels of Fig.~\ref{fig:wavelet} show the temporal evolution of $p$-mode frequency shifts over the period 1996-2012, while 
the lower panels identify periods in which the signal gets 90$\%$  confidence level for both frequency bands. This occurs during periods around the solar maximum (1998-2004) with a significant periodicity of about T = 2 years. 
This shorter cyclic component characterizes different phases of solar magnetic activity, in contrast with activity proxies where the QBP signal is mainly prominent over periods coinciding with solar maximum. This peculiar feature seems to further confirm the persistent nature of the QBP signal, in agreement with previous findings based on the analysis of low degree modes \citep{Sim12}. 
 This finding is important to advance in dynamo theories that already take into account both periodic components of solar magnetic activity. The understanding of the 2 year signal might insert further constraints in these models.
\subsection{Depth and latitudinal dependence}
\begin{figure*}
\begin{center}
\includegraphics[width=5in]{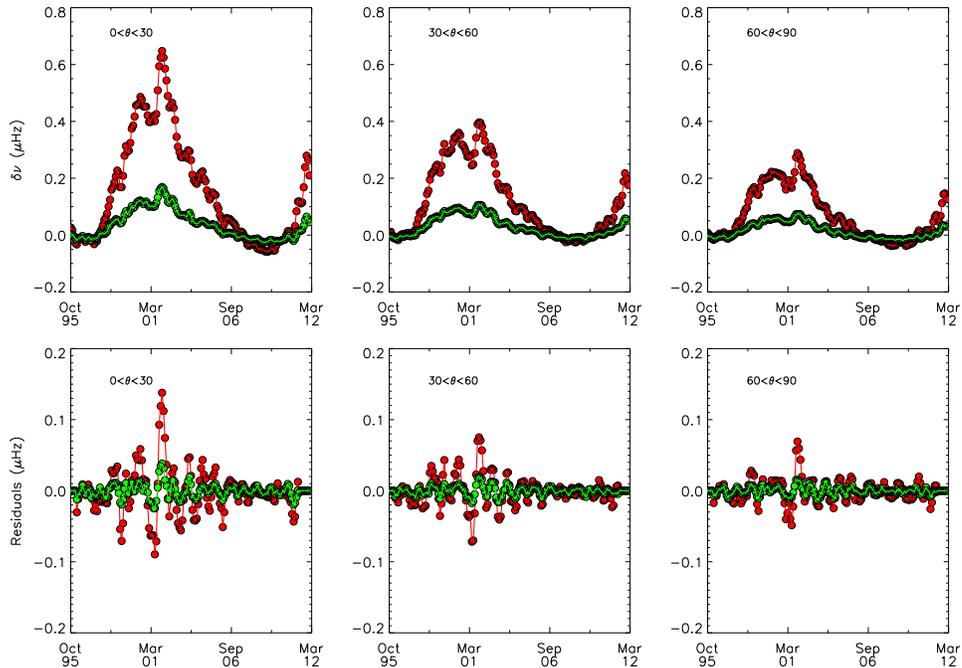}
\caption{The latitudinal dependence of $p$-mode frequency shift in the subsurface layer. The color legend is the same as in Fig.1.} 
\label{fig:2_yr}
\end{center}
\end{figure*}

The $p$-mode spatial configuration can be described by spherical harmonics with each mode characterized by its spherical harmonic degree $\ell$ and azimuthal order $m$. The spherical degree defines the penetration depth through the following relation \citep{Chr91}:
\begin{equation}
r_{t}=\frac{c(r_{t})\sqrt{\ell(\ell+1)}}{2\pi\nu}
\end{equation}
 where c is the sound speed and $r_{t}$ is the LTP radius. A higher value of $\frac{\nu}{\sqrt{\ell(\ell+1)}}$ denotes a smaller value of $r_{t}$ and hence a greater depth. We carried out a depth dependence analysis in four different regions of the Sun's interior (core, radiative zone, tachocline and convection zone) to verify that the shifts we discuss is independently confined. Then
we used the ratio $\frac{\nu}{\sqrt{\ell(\ell+1)}}$ to select only the modes having their LTP between $0.90\le\frac{r_{t}}{R}\le 1$ solar radius. This would verify the frequency dependence of the shift in the subsurface layers and exclude different behavior over the two cycles due, for example, to the presence of a second dynamo mechanism.
Simultaneously we used the ratio $\frac{m}{\ell}$ to select the acoustic modes more sensitive to lower or to higher latitudes. We decided to select three different latitudinal bands corresponding to equatorial latitudes ($0^{0}\le\theta\le 30^{0}$), mid latitudes ($30^{0}\le\theta\le 60^{0}$) and high latitudes ($60^{0}\le\theta\le 90^{0}$).
Some authors have speculated that the QBP signal might be induced by a second dynamo mechanism located at 0.95 solar radius \citep{Ben98a,Ben98b}. It is known that the strong shear present in the subsurface layers increases towards latitudes higher than 60$^{0}$ \citep{Schou98}. If this layer is acting as a second dynamo mechanism and it is causing the QBP, we might find an increases of the 2 yr signal at higher laittudes.
%\subsection{Latitudinal dependence of the shift in the subsurface layers}
 %The frequency shift is a subsurface phenomenon. Nevertheless 
%We carried out a depth dependence analysis in four different regions of the Sun's interior by tuning the ratio $\frac{\nu}{\ell}$ to verify that the shifts we discuss are not influenced by the core, the radiative zone, the tachocline and the convection zone. 
 %Some authors have speculated that the QBP signal might be induced by a second dynamo mechanism located at 0.95 solar radius. It is known that the strong shear present in the subsurface layers increases towards latitudes higher than 60 degree \citep{Schou98}. If this layer is acting as a second dynamo mechanism and it is behind the QBP, we might find an increase of the 2 yr signal at higher latitudes. Therefore we decided to include in our analysis only waves that are confined between 0.90$\le R\le 1$, trying to spot a different latitudinal dependence of the shift possibly caused by this radial shear.
Fig.~\ref{fig:2_yr} shows the changes during the solar cycle 23 as function of latitude for three ranges:% $0\le\theta\le 30^{0}$, $30\le\theta\le 60^{0}$ and above 60$^{0}$.

The figures present several interesting features:
\begin{itemize}
\item the magnitude of the shift decreases with increasing latitudes over both periodic components characterizing solar magnetic activity. This might indicate that the larger magnitude of the signal at equatorial latitudes could be due to the presence of the strong toroidal fields located between -$45^{0}\le\theta\le 45^{0}$ \citep{det00} during the course of the 11 yr cycle;

\item the size of the shift is larger at equatorial latitudes and over periods coinciding with solar maximum at all latitudes and over both periodic components;

\item the 11 yr envelope over the 2 yr cycle is clearly visible at all latitudes.
\end{itemize}
All these features clearly show that the QBP signal is highly coupled with the main dynamo driving the 11 yr cycle.
\section{Origin of the observed frequency shift}
\subsection{Magnitude of the shift versus the subsurface layers}
\begin{figure*}
\begin{center}
\includegraphics[width=4in]{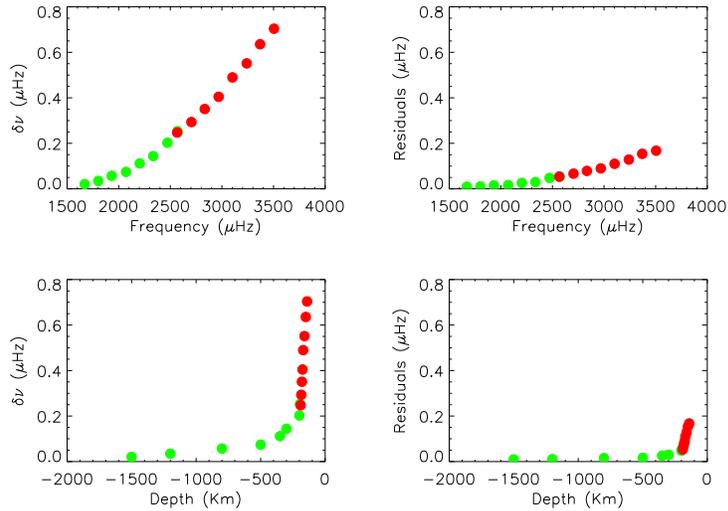}
\caption{The top panels show the frequency dependence of the shifts obtained at the maximum of the solar cycle over the months 08/08/2001-19/10/2001 for the 11 year (left panels) and the 2 year cycles (right panels). The bottom panels show the same amplitudes of the shifts as function of depth from the surface of the Sun. The color legend is the same as in Fig.1.}
\label{fig:dnu_depth}
\end{center}
\end{figure*}

%{\bf We aim to understand if the observed shift caused by the 11 and 2 yr cycle have a common origin. To get the aim we will compare how the magnitude of the shift varies with mode frequency and in the subsurface layers for the two cyclic components.}
The characterization of the shift with frequency, depth and latitude over the 11 and 2 yr cycles has shown differences in the enhancement rate of the magnitude of the shift. We now attempt to provide an explanation for it.
Fig.~\ref{fig:dnu_depth} shows the frequency dependence of the frequency shifts taken over the months 08/08/2001-19/10/2001 (top panels). This period corresponds to the maximum amplitude of the shift for both component of solar activity cycles.
It shows that the enhanced amplitude follows an exponential behavior over solar cycle 23, in agreement with previous findings, and a slower enhancement rate over the 2 yr cycle.
 The bottom panels in Fig.~\ref{fig:dnu_depth} shows the magnitude of the shift as a function of the position of the upper turning point. Those values have been calculated from model S of Christensen Dalsgaard \citep{Cha01}. 
 A sudden increase in the size of the shift is clearly visible in the last few hundred kilometers beneath the solar surface for both periodic components, although it is reduced over the 2 yr period. The acoustic time of the modes is larger in these layers as the sound speed is smaller so any change of magnetic field strength in these layers is amplified.
\subsection{Variation of $\beta$ in the subsurface layers}
The solar magnetic signature along the solar cycle 23 has been extracted from the even-order splitting coefficients of the high degree acoustic modes observed with MDI at respectively 0.996 and 0.999 R$_\odot$. This study shows that poloidal and toroidal magnetic field strengths decrease toward the solar surface \citep{Bal09}.
\begin{figure*}
\begin{center}
\includegraphics[width=5in]{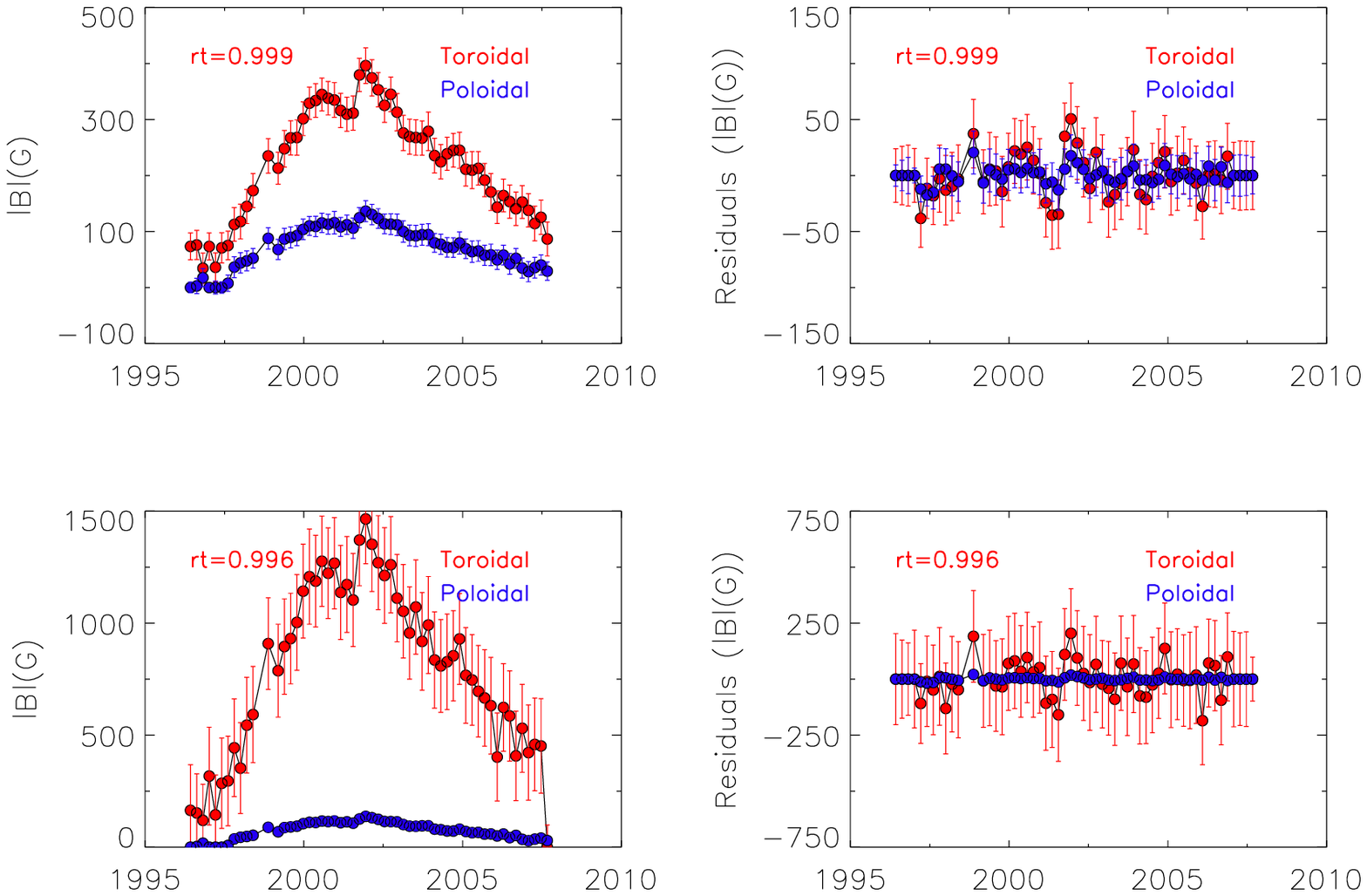}
\caption{Decomposition of the magnetic field strength in poloidal and toroidal fields along solar cycle 23 (left panels) based on \cite{Bal09}. Extraction of the components for the 2 year cycle (right panels) at 0.999 (upper panels) and 0.996 solar radius (lower panels).}
\label{fig:mag_strength}
\end{center}
\end{figure*}
Fig.~\ref{fig:mag_strength} shows the strength of these two magnetic field components along the solar cycle 23 and also deduces the 2 year cycle components at $\frac{r_{t}}{R}=0.999$ and $\frac{r_{t}}{R}=0.996$.
Signatures of the 2 year signal reach $90\%$ of confidence level around the solar maximum (1998-2004) with a period of 2.3 years for both component. This periodicity agrees reasonably well with our findings.
 We now attempt to interpret our findings by taking into account the magnetic field strength behavior in the subsurface layers and the increase of the amplitude  of the modes near the solar surface. We have been misleading if one has deduced that the magnetic field strength should also increase.
 This is not the appropriate interpretation. In the Sun's interior the $\beta=\frac{P_{gas}}{Pmag}>>1$, as the gas pressure is extremely large compared to the magnetic pressure. %Fig.~\ref{fig:beta} shows $\beta$ at solar minimum and solar maximum as function of radius (from 0.996 up to 1 solar radius). 
Only very close to the surface this value becomes smaller \citep{Kos07}. The gas pressure undergoes a strong decay near the surface, while the magnetic pressure, over the same range of depths, undergoes a slower decay (B$^{2}$). As a consequence, $\beta$ reduces dramatically at the vicinity of the surface, so the magnetic pressure plays a direct role on the mode frequency in very close layers near the surface, inciting consequently the observed shift.
%\begin{figure*}
%\begin{center}
%\includegraphics[width=4in]{/usr/local/home/rsimonie/IDLWorkspace81/qbp/PLOTS/beta_surf_min_max.ps}
%\caption{The $\beta$ at solar minimum (left) and maximum (right) as function of solar radius. }
%\label{fig:beta}
%\end{center}
%\end{figure*}
The rapid enhancement in the size of the shift is, therefore, a combination of the $\beta$ ratio and longer acoustic time of the modes in these layers. 
The reduced size of enhacement over the 2 yr cycle may be interpreted in terms of magnetic field strength or topology (see Sec 6.3). 
\subsection{Magnetic field in the subsurface layers from 3D simulation}
The order of magnitude of the magnetic field deduced from splitting values of the sub surface layers \citep{Bal09} give a first insight on what phenomenon can influence the shifts that is discussed in this paper.
 In parallel to this classical approach, dedicated to secular 1D models of the Sun in agreement with deeper helioseismic observations, 3D hydro and magnetohydrodynamic simulations have been developed by Stein \& Nordlund (1998) through the code STAGGER. They  have shown that the order of magnitude of the turbulence pressure represents about 10$\%$ of the gas pressure in the simulations of the subsurface layers of the Sun \citep{Nor00}. Even more interestingly, they have put in evidence  that this turbulent term  concerns a larger region than previously thought, typically at least 1000 km \citep{Ros99}. Magnetohydrodynamical  simulations of 48 Mm large and 20 Mm deep have been performed recently, using an horizontal magnetic field chosen at the bottom of this box compatible with values extracted from \cite{Bal09} and shown in Fig.~\ref{fig:mag_strength}. Two complementary effects are visible in the simulations: a slow reduction of the turbulence by the deeper magnetic field in the lower part of the box and a direct effect of the magnetic field in the last 500 km just below the surface (see also \cite{Piau11}). These simulations agree reasonably with what we observe in the present study, of course the effect is amplified at the near surface in acoustic mode frequencies due to the travel time of the modes. 
\section{Discussion on the physical mechanism behind the QBP}
Seismic signatures of the 11 and 2 yr cycle in $p$-mode frequency shifts might provide a deeper insight on the mechanism behind the cyclic behavior of solar activity. In this section we will interpret our observational findings to shed light on the mechanism behind the 2 yr signal.
\subsection{Second dynamo mechanism}
 In agreement with earlier findings we note that the size of the shift, over the 11 yr cycle, undergoes a sudden and strong enhancement at $\approx$ 200 km below the photosphere, while over the same range of depths we found a reduced gain over the two year cycle. Because of this, some authors invoked the action of a second dynamo mechanism \citep{Fle10, Bro12} due to the subsurface rotational shear extending 5$\%$ below the solar surface \citep{Schou98}. 
 In Sec 3.4 we have shown that the origin and the behavior of the shift for both periodic components of solar magnetic activity in the subsurface layers is due to the sudden and stronger decrease of $\beta$. Therefore we do not visualize a need to invoke a further dynamo mechanism to explain the origin of the shift induced by the 2 yr cycle. Moreover the latitudinal dependence of the size of the shift does not differ over the two cycles and the 11 yr envelope clearly modulates the amplitude of the 2 yr signal at all latitudes, suggesting that the two periodicities are intimately coupled.
 Thus our seismic analysis do not provide observational evidences in favor of a different and separate mechanism from the main dynamo behind the origin of the 2 yr cycle.
 We also want to stress out that our analysis do not exclude the possibility that the strong rotational shear layer might have a role in the generation of toroidal fields \citep{Pip11}, since our analysis indicates that the dynamo independent of its location is behind both component of solar magnetic activity.
\subsection{Instability of magnetic Rossby waves}
The instability of magnetic Rossby waves in the tachocline might enhance the 2 yr periodicity when the magnetic field strength is greater than 10$^{5}$~G \citep{Zaq10}. Such strong magnetic fields are difficult to exist in the tachocline. In fact using splitting of modes with their lower turning points around the base of the convective zone gave an upper limit of 0.3~MG \citep{Bas97}, in good agreement with similar upper values of 0.3-0.4~MG found by other investigators \citep{Ant00}. 
Furthermore it is still an ongoing investigation how the magnetic Rossby waves vary with latitude in slow rotator stars. In order to fit with our results, it should be proven that the Rossby waves are smoothly redistributed along the latitudes with a maximum at low latitude and the poloidal wave number should be such that no nodes will rise from the equator towards the polar regions.
\subsection{Flip-Flop cycle}
%Sunspots are known to preferably appear in a narrow latitudinal belts. Also the longitudinal behavior of sunspot activity shows a noticeable pattern. In fact 
It has been noticed that the biggest active regions tend to appear always at similar longitudes, which are called active longitudes. These are persistent over many cycles, but it can suddenly shift by 180$^{0}$ to the other side of the star \citep{Ber03}. This phenomenon is known as flip-flop cycle \citep{Jet91} and it seems to be rather common in stars \citep{Ber98,Ber02}. 
The origin of the flip-flop cycle is not yet fully understood, but it %is believed to be the result of the beating between a dipole and quadrupole magnetic configuration of the dynamo. In terms of the mean field dynamo theory, stable active longitudes separated by 180$^{0}$ in the Sun and cool active stars 
could be explained as the result of the excitation of a global non axisymmetric (quadrupole) dynamo mode \citep{Mos95,Tuo02,Mos04}. Such dynamo configuration is plausible to exist in the Sun along with the dipole as inferred by solar dynamo models \citep{Mos00}. The relative strengths of the two dynamo modes should define the amplitudes of the observed cycles. Within this formalism it is also predicted that the amplitude of the secondary cycle, in general, is expected to have lower amplitude, as part of the energy is transferred from the primary magnetic configuration (dipolar) to the secondary one (quadrupolar). This feature agrees well with our findings, as we have shown that the secondary cycle in the Sun is an additive contribution to the main cycle, whose signal strength is rather weak.
%\begin{figure}[!b]
%\begin{center}
%\includegraphics[width=2.in]{PICTURES/mag_field_moss.ps}
%\caption{The axisymmetric dipole mode dynamo configuration (dashed line) and the non axisymmetric quadrupole mode dynamo configuration (continuous line) \citep{Mos04}} 
%\label{fig:dynamo_conf}
%\end{center}
%\end{figure}
The period of the oscillations of the axisymmetric mode should, instead, define the lengths of the observed cycles. Therefore the full flip flop cycle is expected to have the same length as the axisymmetric mode \citep{Ber02}. 
In the Sun the major spot activity alternates the active longitudes in about 1.8-1.9 years and on average it has been observed to make 6 switches of the active longitude during the 11 yr sunspot cycle. These values have been obtained by the analysis of solar cycle 18 up to 22 whose lengths were shorter than 11 yr. 
Our findings seems to fit well with predictions and observations of flip-flop cycles. We found a significant periodicity at $\approx$ 2 years, a bit higher than the usual flip-flop findings, but solar cycle 23 lasted longer than usual (about 12.6 years). 
 We, then, might expect that all periodicities between 1.5 yr $\le T\le$4 yr (corresponding to individual periods of one longitude's dominance) might be seen as the visible manifestation of the same physical mechanism. Recently some other authors also concluded that the non-constant period length is the manifestation of a unique quasi-biennial cycle \citep{Vec08,Vec09}. 
 Furthermore the magnitude of the shift increases towards equatorial latitudes over both periodic components of solar magnetic activity and this feature comes up naturally within this formalism.
\section{Conclusion and further perspective}
The analysis of the QBP signal in $\ell$ = 0 to 120 acoustic mode frequency shifts shows that this signal is persistent on the whole solar cycle 23 and the ongoing cycle 24. The significant reduction in the QBP signal strength over the ascending phase of solar cycle 24 
 might be interpreted as signatures of a weaker sub surface magnetic field. The current dynamo models struggle to predict the basic parameters such as duration and strength of the activity cycle and therefore the idea to use acoustic waves as precursor of solar cycle is fascinating. To reach this aim, it will be fundamental to identify the physical mechanism behind it. Our detailed analysis  seems to suggest that the QBP might be the result of the beating between different  dynamo modes.  

Evidences of North-South asymmetry have led some authors to speculate that a significant quadrupolar component is still present \citep{Pul99}, but some others stated that since the end of the Maunder Minimum the solar field has been dipolar \citep{Tob96,Tob98}. If the QBP signal is confirmed to be the result of the beating between different dynamo modes, then this work provides further evidence in favor of the existence of a quadrupolar component in the Sun.%, as signatures of the 2 yr cycles have been found in solar activity proxies since 1853. 
It will be extremely important, therefore, to look for signatures of the quadrupolar magnetic dynamo  configuration. It is not a simple task, but within this context the investigation of the seismic properties of the QBP signal might play a key role. 
 For example some authors found the periodicity from the northern hemisphere to differ from the one in the south hemisphere \citep{Vec08,Vec09,Ber03}. It will be worth to check if the analysis of $p$-mode frequency shift or other $p$-mode parameters in the two hemisphere might also spot asymmetries. Solar cycle 24 could be a good candidate to look for asymmetry. Infact we have already shown the signatures of the 2 yr cycle over its ascending phase.
Furthermore it will be equally important to investigate magnetic activity on other solar type stars using long term data provided, for example, by asteroseismology missions (such as COROT, Kepler). The availability of a sample of solar analogues selected at different stages of their evolution will give us the opportunity to infer accurate relations between activity quantities, such as the length and amplitude of cycles and stellar properties such as rotation, age and convective envelope depth. Solar and stellar activity requires more systematic and detailed studies to advance, observationally and theoretically, our understanding of the dynamo.
\acknowledgments
This work utilizes GONG data obtained by the NSO
Integrated Synoptic Program (NISP), managed by the
National Solar Observatory, which is operated by AURA,
Inc. under a cooperative agreement with the National
Science Foundation.  The data were acquired by instruments
operated by the Big Bear Solar Observatory, High Altitude
Observatory, Learmonth Solar Observatory, Udaipur Solar
Observatory, Instituto de Astrofisica de Canarias, and
Cerro Tololo Interamerican Observatory.
\bibliographystyle{apj}
\bibliography{biblio1}

\begin{thebibliography}{55}
\expandafter\ifx\csname natexlab\endcsname\relax\def\natexlab#1{#1}\fi

\bibitem[{{Anderson.} {et~al.}(1990){Anderson.}, {Duvall}, \&
  {Jefferies}}]{And90}
{Anderson.}, E.~R., {Duvall}, T. L.~J., \& {Jefferies}, S.~M. 1990, ApJ, 364,
  699

\bibitem[{{Antia} {et~al.}(2000){Antia}, {Chitre}, \& {Thompson}}]{Ant00}
{Antia}, H.~M., {Chitre}, S., \& {Thompson}, M. 2000, \aap, 360, 335

\bibitem[{{Bai}(2003)}]{Bai03}
{Bai}, T. 2003, ApJ, 585, 1114

\bibitem[{{Baldner} {et~al.}(2009){Baldner}, {Antia}, {Basu}, \&
  {Larson}}]{Bal09}
{Baldner}, C., {Antia}, H., {Basu}, S., \& {Larson}, T.~M. 2009, \aap, 705,
  1704

\bibitem[{{Baliunas} {et~al.}(1995){Baliunas}, {Donahue}, {Soon},
  {et~al.}}]{Bali95}
{Baliunas}, S.~L., {Donahue}, R.~A., {Soon}, W.~H., {et~al.} 1995, \apj, 438,
  269

\bibitem[{{Basu}(1997)}]{Bas97}
{Basu}, S. 1997, MNRAS, 288, 572

\bibitem[{{Benevolenskaja}(1998{\natexlab{a}})}]{Ben98a}
{Benevolenskaja}, E. 1998{\natexlab{a}}, ApJ, 509L, 49

\bibitem[{{Benevolenskaja}(1998{\natexlab{b}})}]{Ben98b}
---. 1998{\natexlab{b}}, Sol.Phys, 181, 479

\bibitem[{{Berdyugina} \& {Tuominen}(1998)}]{Ber98}
{Berdyugina}, S.~V., \& {Tuominen}, I. 1998, \aap, 336, 25

\bibitem[{{Berdyugina} \& {Tuominen}(2002)}]{Ber02}
---. 2002, \aap, 394, 505

\bibitem[{{Berdyugina} \& {Usoskin}(2003)}]{Ber03}
{Berdyugina}, S.~V., \& {Usoskin}, I.~G. 2003, \aap, 405, 1121B

\bibitem[{{B\"{o}hm-Vitense}(2007)}]{Boo07}
{B\"{o}hm-Vitense}, E. 2007, \apj, 657, 486

\bibitem[{{Brandenburg} {et~al.}(1998){Brandenburg}, {Saar}, \&
  {Turpin}}]{Bra98}
{Brandenburg}, A., {Saar}, S.~H., \& {Turpin}, C.~J. 1998, \apj, 498, L51

\bibitem[{{Broomhall} {et~al.}(2009){Broomhall}, {Chaplin}, {Elsworth},
  {Fletcher}, \& {New}}]{Bro09a}
{Broomhall}, A.~M., {Chaplin}, W.~J., {Elsworth}, Y., {Fletcher}, S.~T., \&
  {New}, R. 2009, \apj, 700L

\bibitem[{{Broomhall} {et~al.}(2012){Broomhall}, {Chaplin}, {Elsworth}, \&
  {Simoniello}}]{Bro12}
{Broomhall}, A.~M., {Chaplin}, W.~J., {Elsworth}, Y., \& {Simoniello}, R. 2012,
  MNRAS, 420

\bibitem[{{Chaplin} {et~al.}(2001){Chaplin}, {Appourchaux}, \& {Isaak}}]{Cha01}
{Chaplin}, W.~J., {Appourchaux}, T.~{Elsworth}, Y.~P., \& {Isaak}, G.~R. 2001,
  MNRAS, 324, 910

\bibitem[{{Christensen-Dalsgaard} \& {Berthomieu}(1991)}]{Chr91}
{Christensen-Dalsgaard}, J., \& {Berthomieu}, J. 1991, in Solar Interior and
  atmosphere, ed. A.N. Cox, W.C. Livingstone and M.Matthews (Tucson: University
  of arizona Press), 401

\bibitem[{{de Toma} {et~al.}(2000){de Toma}, {White}, \& {Harvey}}]{det00}
{de Toma}, G., {White}, O.~R., \& {Harvey}, K.~L. 2000, \apj, 529, 1101

\bibitem[{{Fletcher} {et~al.}(2010){Fletcher}, {Broomhall}, {Salabert},
  {et~al.}}]{Fle10}
{Fletcher}, S., {Broomhall}, A.~M., {Salabert}, D., {et~al.} 2010, \apj, 718L,
  19

\bibitem[{{Fluri} \& {Berdyugina}(2004)}]{Flu04}
{Fluri}, D.~M., \& {Berdyugina}, S. 2004, Sol.Phys., 224, 153

\bibitem[{{Garc\'ia} {et~al.}(2010){Garc\'ia}, {Mathur}, {Salabert},
  {et~al.}}]{Gar10}
{Garc\'ia}, R.~A., {Mathur}, S., {Salabert}, D., {et~al.} 2010, Science, 329,
  1032

\bibitem[{{Hill} {et~al.}(1998){Hill}, {Anderson}, {Howe}, {et~al.}}]{Hil98}
{Hill}, F., {Anderson}, E., {Howe}, R., {et~al.} 1998, ESASP, 231

\bibitem[{{Howe} {et~al.}(2002){Howe}, {Komm}, \& {Hill}}]{How02}
{Howe}, R., {Komm}, R., \& {Hill}, F. 2002, \apj, 580, 1172

\bibitem[{{Jain} {et~al.}(2001){Jain}, {Tripathy}, \& {Bhatnagar}}]{Jai01}
{Jain}, K., {Tripathy}, S.~C., \& {Bhatnagar}, A. 2001, \apj, 542, 521

\bibitem[{{Jain} {et~al.}(2011){Jain}, {Tripathy}, \& {Hill}}]{Jai11}
{Jain}, K., {Tripathy}, S.~C., \& {Hill}, F. 2011, \apj, 739, 6

\bibitem[{{Jetsu} {et~al.}(1991){Jetsu}, {Pelt}, {Tuominen}, \&
  {Nations}}]{Jet91}
{Jetsu}, L., {Pelt}, J., {Tuominen}, I., \& {Nations}, H. 1991, Proc. IAU Coll.
  130, The Sun and Cool stars; Activity, Magnetism, Dynamics, Springer-Verlag
  Heidelberg, 381

\bibitem[{{Koll\'ath} \& {Ol\'ah}(2009)}]{Kol09}
{Koll\'ath}, Z., \& {Ol\'ah}, K. 2009, AA, 501, 695

\bibitem[{{Kosovichev}(2007)}]{Kos07}
{Kosovichev}, A.~G. 2007, AdSpR, 41, 380

\bibitem[{{Krivova} \& {Solanki}(2002)}]{Kri02}
{Krivova}, N.~A., \& {Solanki}, S.~K. 2002, AA, 394, 701

\bibitem[{{Messina} \& {Guinan}(2002)}]{Mes02}
{Messina}, S., \& {Guinan}, E.~F. 2002, AA, 393, 225

\bibitem[{{Metcalfe} {et~al.}(2010){Metcalfe}, {Basu}, {Henry},
  {et~al.}}]{Met10}
{Metcalfe}, T.~S., {Basu}, S., {Henry}, T.~J., {et~al.} 2010, A$\&$A, 723L, 213

\bibitem[{{Moss}(2004)}]{Mos04}
{Moss}, D. 2004, MNRAS, 352, 17

\bibitem[{{Moss}(2008)}]{Mos08}
---. 2008, MNRAS, 306, 300

\bibitem[{{Moss} {et~al.}(1995){Moss}, {Barker}, {Brandenburg}, \&
  {Tuominen}}]{Mos95}
{Moss}, D., {Barker}, D., {Brandenburg}, A., \& {Tuominen}, I. 1995, \aap, 294,
  155

\bibitem[{{Moss} \& {Brooke}(2000)}]{Mos00}
{Moss}, D., \& {Brooke}, J. 2000, MNRAS, 315, 521

\bibitem[{{Nordlund} \& {Stein}(2000)}]{Nor00}
{Nordlund}, A., \& {Stein}, R.~F. 2000, ASPCS, 203, 362

\bibitem[{{Ol\'ah} \& {Strassmeier}(2002)}]{Ola02}
{Ol\'ah}, K., \& {Strassmeier}, K.~G. 2002, AN, 323, 361

\bibitem[{{Piau} {et~al.}(2011){Piau}, {Stein}, {Melo}, {et~al.}}]{Piau11}
{Piau}, L., {Stein}, R.~F., {Melo}, S., {et~al.} 2011, SF2A, eds G. Alecian et
  al.

\bibitem[{{Pipin} \& {Kosovichev}(2011)}]{Pip11}
{Pipin}, V.~V., \& {Kosovichev}, A.~G. 2011, \apj, 104, 1

\bibitem[{{Pulkkinen} {et~al.}(1999){Pulkkinen}, {Baker}, {Cogger},
  {et~al.}}]{Pul99}
{Pulkkinen}, T.~I., {Baker}, D.~N., {Cogger}, L.~L., {et~al.} 1999, JGR, 14,
  10284

\bibitem[{{Ritzwoller} \& {Lavely}(1991)}]{Rit91}
{Ritzwoller}, M.~H., \& {Lavely}, E.~M. 1991, ApJ, 369, 557

\bibitem[{{Rosenthal} {et~al.}(1999){Rosenthal}, {Christensen-Dlasgaard},
  {Nordlund}, {Stein}, \& {Trampedach}}]{Ros99}
{Rosenthal}, C.~S., {Christensen-Dlasgaard}, J., {Nordlund}, A., {Stein},
  R.~F., \& {Trampedach}, R. 1999, A$\&$A, 351, 689

\bibitem[{{Salabert} {et~al.}(2010){Salabert}, {Garc\'ia}, {Pall\'e}, \&
  {Jim\'enez-Reyes}}]{Sal10}
{Salabert}, D., {Garc\'ia}, R.~A., {Pall\'e}, P.~L., \& {Jim\'enez-Reyes},
  S.~J. 2010, ASPC, 428, 51S

\bibitem[{{Schou} {et~al.}(1998){Schou}, {Antia}, {Basu}, {et~al.}}]{Schou98}
{Schou}, J., {Antia}, H.~M., {Basu}, S., {et~al.} 1998, ApJ, 505, 390

\bibitem[{{Simoniello} {et~al.}(2010){Simoniello}, {Finsterle}, {Garc\'ia},
  {et~al.}}]{Sim10}
{Simoniello}, R., {Finsterle}, W., {Garc\'ia}, R.~A., {et~al.} 2010, A$\&$A,
  516, 30

\bibitem[{{Simoniello} {et~al.}(2012){Simoniello}, {Finsterle}, {Salabert},
  {et~al.}}]{Sim12}
{Simoniello}, R., {Finsterle}, W., {Salabert}, D., {et~al.} 2012, A$\&$A, 539,
  135

\bibitem[{{Tobias}(1996)}]{Tob96}
{Tobias}, S.~M. 1996, A$\&$A, 307, L21

\bibitem[{{Tobias}(1998)}]{Tob98}
---. 1998, MNRAS, 296, 653

\bibitem[{{Torrence} \& {Compo}(1998)}]{Tor98}
{Torrence}, C., \& {Compo}, G. 1998, Bull.Am.Met.Soc, 79, 61

\bibitem[{{Tuominen} {et~al.}(2002){Tuominen}, {Berdyugina}, \&
  {Korpi}}]{Tuo02}
{Tuominen}, I., {Berdyugina}, S., \& {Korpi}, M. 2002, AN, 323, 367

\bibitem[{{Usoskin} {et~al.}(2007){Usoskin}, {Solanki}, \& {Kovaltsov}}]{Uso07}
{Usoskin}, I., {Solanki}, S., \& {Kovaltsov}, G.~A. 2007, A$\&$A, 471, 301

\bibitem[{{Vecchio} \& {Carbone}(2008)}]{Vec08}
{Vecchio}, A., \& {Carbone}, V. 2008, ApJ, 683, 536

\bibitem[{{Vecchio} {et~al.}(2009){Vecchio}, {Cauzzi}, \& {Reardon}}]{Vec09}
{Vecchio}, A., {Cauzzi}, G., \& {Reardon}, K. 2009, A$\&$A, 494, 629

\bibitem[{{Zaqarashvili} {et~al.}(2010){Zaqarashvili}, {Carbonell}, {Oliver},
  \& {Ballester}}]{Zaq10}
{Zaqarashvili}, T., {Carbonell}, M., {Oliver}, R., \& {Ballester}, J.~L. 2010,
  ApJL, 724, 95

\bibitem[{{Zaqarashvili} {et~al.}(2011){Zaqarashvili}, {Oliver}, {Ballester},
  {Carbonell}, {Khodachenko}, {et~al.}}]{Zaq11}
{Zaqarashvili}, T., {Oliver}, R., {Ballester}, J.~L., {Carbonell}, M.,
  {Khodachenko}, M.~L., {et~al.} 2011, A$\&$A, 532, 139

\end{thebibliography}

\end{document}